\documentclass[a4paper,11pt]{article}
\usepackage{aaskaiid}
\usepackage{hyperref}
\usepackage{orcidlink}
\usepackage{longtable}

\title{Very Long Baseline Interferometry with the SKA Observatory}
\ShortTitle{VLBI with the SKAO}
\ShortName{Radcliffe, Yang \& the VLBI Science Working Group} 

\author[1,2,3]{Jack~F.~Radcliffe\orcidlink{0000-0002-0813-0497}}
\author[4]{Jun~Yang}
\author[5]{Zsolt~Paragi}
\author[6]{Olga~Bayandina\orcidlink{0000-0003-4116-4426}}
\author[3]{Cormac~Reynolds}
\author[8]{Cristina~García-Miró}
\author[7]{Tao~An}

\affiliation[1]{Jodrell Bank Centre for Astrophysics, University of Manchester, Oxford Road, Manchester M13 9PL, UK}
\affiliation[2]{Curtin Institute of Radio Astronomy (CIRA), Curtin University, 1 Turner Avenue, Technology Park, Bentley, WA 6102, Australia}
\affiliation[3]{CSIRO Space and Astronomy, PO Box 1130, Bentley, WA 6102, Australia}
\affiliation[4]{Department of Physics and Astronomy, Chalmers University of Technology, Onsala Space Observatory, SE-439 92 Onsala, Sweden}
\affiliation[5]{JIVE (Joint Institute for VLBI ERIC), Oude Hoogeveensedijk 4, 7991 PD Dwingeloo, The Netherlands}
\affiliation[6]{SKA Observatory, 2 Fir Street, Black River Park, Observatory, 7925, Cape Town, South Africa (Olga Bayandina)}
\affiliation[7]{Department of Astronomy, University of Science and Technology of China, 96 Jinzhai Road, Hefei, Anhui 230026, China}
\affiliation[8]{Observatorio de Yebes, Observatorio Astron\'omico Nacional (IGN), Cerro de la Palera s/n, 19141 Yebes, Guadalajara, Spain}

\renewcommand{\hookAfterAffiliation}{%
  \vskip4pt
  \noindent\small\itshape on behalf of the VLBI Science Working Group --- the complete list of authors and affiliations is given in Appendix~\ref{app:fullauthorlist}.%
  \normalfont%
}

\emailAdd{jack.radcliffe@manchester.ac.uk} \emailAdd{jun.yang@chalmers.se}

\abstract{Very Long Baseline Interferometry (VLBI) is a fundamental technique that combines signals from telescopes separated by thousands of kilometres to obtain some of the highest-resolution images in astronomy. The Square Kilometre Array Observatory (SKAO) will provide sensitive elements to current VLBI arrays, enabling deep VLBI observations (including the first low-frequency VLBI capability in the Southern Hemisphere). This summary chapter outlines the science that the SKAO, when participating in a VLBI array, will achieve (such as AGN physics, explosive transient localisation, Galactic structure, and gravitational waves), and the current understanding of how VLBI with the SKAO will operate. Given that VLBI requires partnerships with other instruments, the chapter concludes with outstanding issues and considerations for the community needed to maximise the scientific return on investment in SKA-VLBI.}

\begin{document}
\maketitle





\section{Introduction}

Very Long Baseline Interferometry (VLBI) has been a cornerstone technique in radio astronomy for more than half a century, delivering the highest angular resolution observations in astrophysics \citep[e.g.,][]{Thompson2017isra}. By correlating signals from widely separated radio antennas across the globe and in space, VLBI achieves angular resolutions as fine as tens of microarcseconds at centi- and millimetre wavelengths, enabling science that probes some of the most extreme environments in the Universe. Landmark achievements include directly imaging the shadows of supermassive black holes \citep[e.g.,][]{eht2019L1}, constraining jet-launching regions in active galactic nuclei (AGN) \citep[e.g.,][]{2023Natur.616..686L}, and providing the astrometric precision that underpins the International Celestial Reference Frame \citep[e.g.,][]{Charlot2020AandA}. VLBI has also proven a powerful astrophysical probe beyond continuum imaging: spectral line observations of Galactic masers have mapped the three-dimensional kinematic structure of the Milky Way \citep[e.g.,][]{Brunthaler2011AN}, while megamaser disks around AGN provide geometric distance measurements that yield some of the most precise, single-step determinations of the Hubble constant \citep[e.g.,][]{Pesce2020ApJL}. Pulsar astrometry with VLBI delivers microarcsecond-precision distances and proper motions that test theories of gravity and constrain neutron star populations \citep[e.g.,][]{Deller2016ApJ}. Outside of astronomy, VLBI plays a fundamental and societally important role in spacecraft tracking \citep[][and references therein]{Gurvits+2023SSR} and geodesy -- the study of the Earth's dynamic shape, size and gravity \citep{Petit2010IERS,Nothnagel2017JGeo}.

The Square Kilometre Array Observatory (SKAO) will transform VLBI capabilities. Both SKA-Mid in South Africa and SKA-Low in Western Australia will support phased-array beamforming, allowing them to operate as ultra-sensitive elements in global VLBI networks \citep{Dewdney2009SKA,Paragi2015AASKA}. When phased up, SKA-Mid, operating at frequencies between 0.35--15.4\,GHz, will offer a collecting area comparable to the largest existing radio telescopes, dramatically improving the sensitivity and imaging fidelity of current VLBI arrays. Its digital architecture will enable simultaneous participation in large-scale surveys and targeted long-baseline observations. SKA-Low, operating at frequencies below 350\,MHz, will pioneer low-frequency VLBI in the Southern Hemisphere, opening a new parameter space for studies of, e.g., cosmic magnetism, galaxy evolution, pulsars, and ionospheric science.

\begin{figure}[htp]
    \centering
    \includegraphics[width=\linewidth]{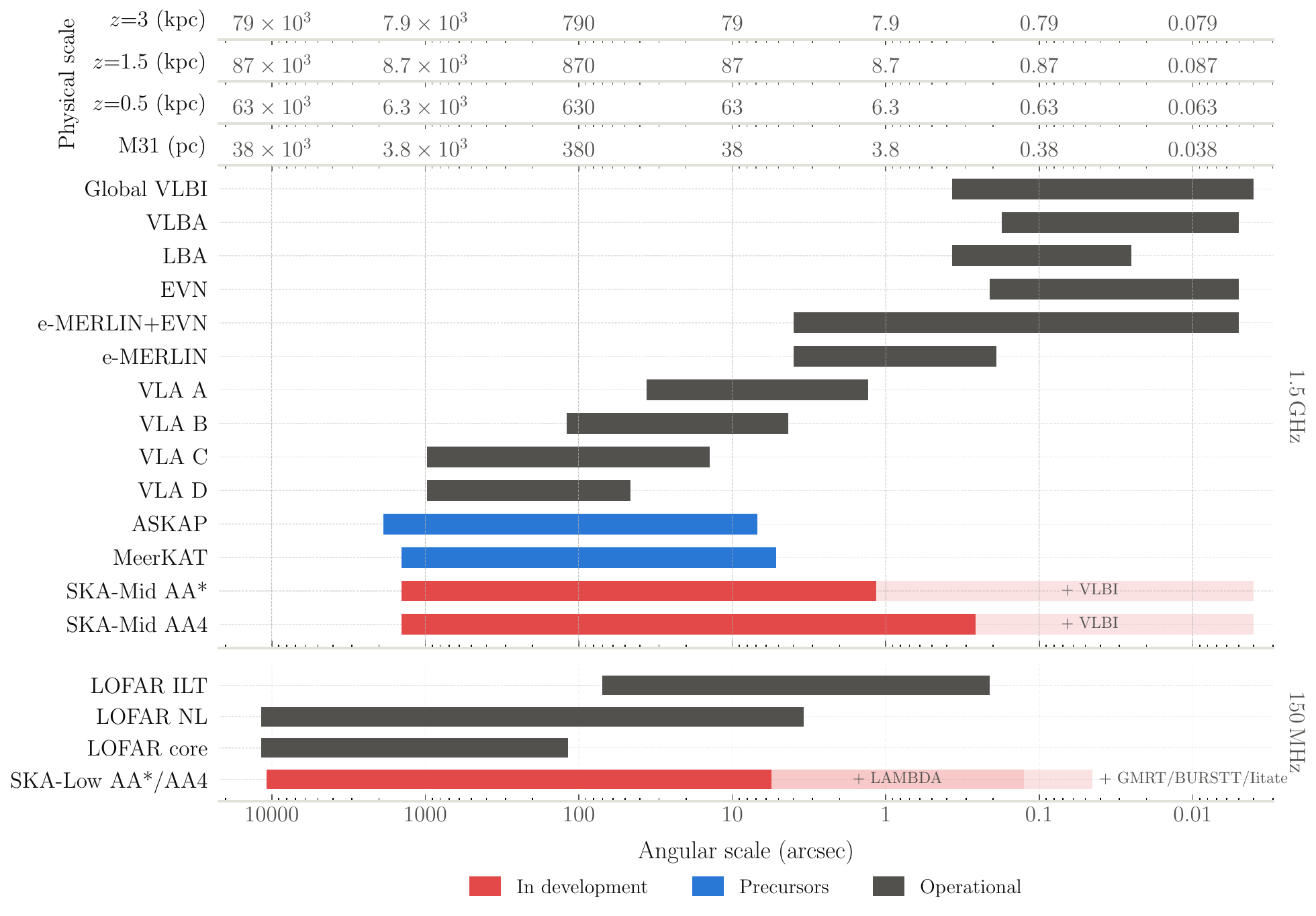}
    \caption{The range of angular scales probed by a selection of existing radio interferometric instruments, and the scales probed by the SKAO. The top panel shows SKA-Mid at AA$^*$ and AA4 at a central frequency of 1.5\,GHz, including the SKAO precursors MeerKAT and ASKAP, while the bottom panel shows SKA-Low at AA$^*$/AA4 at a central frequency of 150\,MHz, including the core, Dutch (NL) and international (ILT) configurations of LOFAR. Both panels are drawn on the same angular scale, with all scales taken as $\lambda/B$ from the shortest and longest baselines of each array, and the upper axis gives the corresponding proper physical scale at a range of distances (M31/780 kpc, and redshifts of 0.5, 1.5 and 3). Neither SKA telescope will be able to probe the milliarcsecond regime without the added VLBI capability. The pale extensions show the resolutions reached by adding VLBI partner stations: for SKA-Low, these include the prospective LAMBDA stations \citep[see][]{Timmerman01.2026.SKA} and then intercontinental baselines to the GMRT, BURSTT and Iitate.}
    \label{fig:las_skavlbi}
\end{figure}

VLBI will also transform the capabilities of the SKAO by providing the long baselines to probe astrophysical phenomena at the sub-arcsecond to milliarcsecond scale, something that the SKAO will not be able to do alone (see Figure~\ref{fig:las_skavlbi}). Imaging of AGN and relativistic jets at milliarcsecond scales will reveal jet-launching and collimation mechanisms, magnetic-field structure, and energy-dissipation processes across cosmic time \citep[e.g.,][]{Zensus1997ARA,Blandford2019ARA}. Precision astrometry, enabled by SKAO's multi-beam capabilities, will reach new frontiers, allowing accurate distance and proper-motion measurements for pulsars and maser sources throughout the Galaxy \citep[e.g.,][]{Reid2014ARA}. These measurements will refine Galactic structure and rotation, measure black hole masses, and help detect gravitational waves by measuring pulsar parallaxes to high precision.

In the time domain, the ability to rapidly localise fast radio bursts (FRBs) and other fast transients with sub-milliarcsecond accuracy will provide critical insight into their progenitors and host environments \citep[e.g.,][]{Chatterjee2017Nature,Marcote2017ApJL,Kirsten2022Nature}. Tidal disruption events, core-collapse supernovae, and gamma-ray burst afterglows will also be observed at the highest resolutions, tracing shock and jet evolution and energy injection on parsec scales.

Beyond astrophysics, SKA-VLBI will strengthen global geodesy and navigation. Its Southern Hemisphere location and frequency coverage will improve the geometric distribution and sensitivity of geodetic VLBI networks, improving the stability of terrestrial and celestial reference frames and supporting planetary missions. Dedicated surveys of the Southern sky will reach sources that are inaccessible to Northern Hemisphere arrays, extending the celestial reference frame southwards and densifying the catalogue of compact calibrators on which differential astrometry depends \citep{MingHuiXu01.2026.SKA}. SKA-VLBI may also contribute to fundamental physics by constraining Lorentz invariance, testing General Relativity in the strong-field regime, and probing gravitational wave backgrounds through high-precision pulsar timing and astrometric microlensing studies \citep{Gwinn1997ApJ,Manchester2013CQGra,Deller2019Sci}.

This chapter provides an overview of SKA-VLBI capabilities and the science they enable. In Section~\ref{s:science}, we summarise the main science themes emerging from the VLBI-focused chapters of this book, and the small set of capabilities that almost all of them require. In Section~\ref{s:technical}, we describe how VLBI will be implemented at the SKAO, covering tied-array beamforming and data output, the dedicated data path to an external correlator, and the staged delivery of the capability through science verification to Cycle~0. In Section~\ref{s:challenges}, we set out the principal challenges that remain -- from the need for a coordinating body for global VLBI participation, through the bandwidth and image fidelity limitations imposed by current partner arrays, to the shortage of long-baseline stations, particularly for SKA-Low -- together with our recommendations to the wider community. We present our conclusions in Section~\ref{s:conclusions}.

\section{Science with SKA-VLBI}\label{s:science}

Across all the VLBI-focused chapters, a common theme runs through them: many astrophysical questions require compact structures to be resolved or small motions to be measured, and current arrays can do neither at the required sensitivities. SKA-VLBI addresses this on three fronts at once, combining sub-milliarcsecond imaging with microarcsecond astrometry and the ability to respond on transient timescales, all at microJansky ($\mu$Jy) sensitivities. It also brings these capabilities to the Southern Hemisphere, where high-sensitivity VLBI has been scarce and where much of the accessible SKAO sky lies. VLBI is therefore better understood as a \textbf{core SKAO capability} than as an addition, since it provides the link between small-scale physics and the large-scale processes the observatory is built to study.

The chapters consistently highlight VLBI's role in multi-messenger surveys. Coordinated efforts across optical, X-ray, gamma-ray, gravitational-wave, and neutrino observatories rely more on high-resolution data, including proper motions, shock structures, and jet morphology. Without accompanying high-sensitivity and high-resolution VLBI observations, transient alerts and long-term monitoring are difficult to interpret. This point is especially true for transients and accreting compact objects. In these cases, the radio band traces non-thermal particle populations and magnetic field structure in a way no other band does, and allows us to constrain source geometry, which breaks degeneracies that otherwise leave the physical parameters governing the emission of the ejecta poorly constrained \citep[e.g.,][]{Shu01.2026.SKA,Giarratana01.2026.SKA,TaoAn01.2026.SKA}.

We summarise the main scientific themes below, and refer the reader to the individual chapters for the details.

\subsection{Key science themes}

\paragraph{Accreting black holes and relativistic jets.}
This theme accounts for more VLBI chapters than any other. How jets are launched, collimated, and ultimately dissipated remains poorly understood, and will stay that way until the flow can be imaged directly across the full range of scales involved, from the innermost accretion flow out to the kiloparsec-scale feedback regions \citep{Baczko01.2026.SKA,Pathak01.2026.SKA}. The SKAO with a VLBI capability, together with higher-frequency VLBI observations, will span this range of spatial scales, as shown in Figure~\ref{fig:las_skavlbi}.

At high redshift ($z>0.5$), VLBI population studies will probe black-hole growth and galaxy formation, provided that the first powerful radio sources can be resolved and characterised individually \citep{Spingola01.2026.SKA}. At lower redshift, the emerging class of high-energy-emitting radio galaxies will provide laboratories for misaligned jets and for particle acceleration operating across several spatial scales simultaneously \citep{Bruni01.2026.SKA}. Compact FR0 systems and radio-quiet AGN pose the opposite problem: their cores are weak, and their outflows carry little power, so detecting them at all will require the sensitivity that the SKAO provides \citep{Baldi01.2026.SKA,Panessa01.2026.SKA}. Neutral hydrogen absorption offers a complementary view of the same feeding and feedback cycle, as milliarcsecond mapping will resolve the kinematics and morphology of the circumnuclear gas \citep{Pandey-Pommier03.2026.SKA}.

\paragraph{Compact binaries and stellar endpoints.}
In X-ray binaries, the inflow-outflow coupling evolves over hours to days, so resolving the jets and their polarisation will require sensitive VLBI observations, with improved $uv$ coverage, at a correspondingly high cadence \citep{TaoAn01.2026.SKA}. Colliding-wind and gamma-ray binaries need similar resolution to resolve their shock structures and locate the non-thermal particles, and the same applies to surveying the population rather than revisiting a few well-studied systems \citep{Marcote01.2026.SKA}. Magnetar proper motions will test formation scenarios by tying individual objects back to their parent supernova remnants \citep{Akahori02.2026.SKA}, while precise orbital measurements of black-hole-neutron-star binaries and intermediate-mass black-hole candidates will constrain both strong-field gravity and the evolutionary channels that produce such systems \citep{Lin01.2026.SKA}.

\paragraph{Time-domain explosive transients.}
For fast transients, the key factors are the SKAO's quick responsiveness and VLBI's ability to measure ejecta sizes directly. Milliarcsecond imaging of tidal disruption events will distinguish a collimated jet from a wide-angle wind, constrain the magnetic field structure, and establish whether the emission is genuinely nuclear \citep{Shu01.2026.SKA}. Gamma-ray burst afterglows must be observed while they are still expanding if the measurements are to discriminate between jet geometries and circumburst environments \citep{Giarratana01.2026.SKA}. For supernovae, VLBI remains the only practical way to resolve the ejecta at early times and to connect progenitor properties to the outcome of the explosion \citep{TaoAn02.2026.SKA}.

\paragraph{Precision astrometry and Galactic structure.}
Astrometry serves as both a core scientific focus and a foundational capability for other research areas. In the southern sky, astrometric and geodetic surveys will enhance the celestial reference frame, especially in regions where it is currently weakest\citep{MingHuiXu01.2026.SKA}. Maser astrometry will enable the creation of a three-dimensional map of spiral structures in the Milky Way, and place constraints on the Galactic bar and overall Galactic parameters \citep{XuYe01.2026.SKA}. Surveys of the Galactic Centre using large OH-maser proper-motion observations will reveal the spatial and kinematic properties of the nuclear stellar disk  \citep{Imai01.2026.SKA}. Achieving microarcsecond accuracy in such large surveys will rely heavily on calibration strategies, along with sensitivity, with multi-beam and MultiView techniques making this routine \citep{Rioja01.2026.SKA}. This level of precision will also allow the detection of planetary companions and the radio emissions from low-mass stars and ultra-cool dwarfs, providing insights into their magnetospheres \citep{Curiel01.2026.SKA}.

\paragraph{Fundamental physics and gravitational-waves.}
This precision extends to tests of gravity as well. Measuring how Solar System bodies bend light will improve constraints on relativistic parameters and, over time, on higher-order gravitational effects \citep{Li01.2026.SKA}. Pulsar timing arrays depend on precise measurements of distances and positions to interpret the nanohertz gravitational-wave signal, and accurate localisation is especially the limiting factor for continuous-wave sources \citep{Takahashi01.2026.SKA}. Electromagnetic counterparts of dual and binary supermassive black holes form part of the same gravitational background, and a systematic radio census would link these two views of the population for the first time \citep{DAmato01.2026.SKA}. Large single-dish facilities will also play a role by conducting timing programs that increase the sensitivity of pulsar timing arrays \citep{Chen01.2026.SKA}.

\paragraph{Low-frequency VLBI.}
Low-frequency VLBI remains the least developed among these areas. Precise low-frequency source catalogues are essential for Epoch of Reionisation experiments, as inaccuracies in foreground-source positions can persist in residuals after subtraction \citep{Kobayashi01.2026.SKA}. Longer baselines covering 50--350 MHz will also provide the resolution needed for steep-spectrum sources, pulsar astrometry, scintillometry, and the low-frequency counterparts of transients \citep{Timmerman01.2026.SKA}. Achieving this requires technological advancements alongside the scientific efforts, pointing toward next-generation extensions of SKA-Low such as LAMBDA.

\paragraph{Infrastructure, and technical developments.}
All these scientific cases rely on parallel technological advancements. Space-VLBI will enhance angular resolution and $uv$ coverage beyond what any ground-based array can achieve, with SKA-Mid serving as the sensitive core of such a network and southern targets gaining the most \citep{Kovalev01.2026.SKA}. On Earth, new stations across Africa will provide the north-south baselines that the southern array currently lacks, directly improving image fidelity \citep{Bempong-Manful01.2026.SKA}. Additionally, new large dishes and demonstrator facilities will boost sensitivity, scheduling flexibility, and contribute to science in their own right \citep{Wongphecauxson01.2026.SKA,Chen01.2026.SKA}. Despite this, $uv$ coverage will remain sparse, requiring ongoing improvements in imaging and deconvolution to translate sensitivity into improved dynamic range \citep{LeZhang01.2026.SKA}. Interoperability with EVN, LBA, VLBA, and ngVLA will also influence how much scientific progress is realistically achievable \citep{Kadler02.2026.SKA}.

\subsection{SKA-VLBI drivers}
To finish this section, we summarise the overall set of capabilities required by almost every one of these science cases:
\begin{itemize}
\item Milliarcsecond to microarcsecond angular resolution for resolving compact jets, shocks, and binaries \citep[e.g.,][]{Baczko01.2026.SKA,Spingola01.2026.SKA,DAmato01.2026.SKA}.
\item $\mu$Jy-level sensitivity and polarimetry to detect weak non-thermal components and constrain magnetic fields \citep[e.g.,][]{Kadler01.2026.SKA,Shu01.2026.SKA}.
\item High-cadence, rapid-trigger VLBI for transient evolution on hours-to-days timescales \citep[e.g.,][]{Giarratana01.2026.SKA,TaoAn01.2026.SKA}.
\item Wide-band, multi-beam tied-array operation and MultiView calibration to suppress systematics and enable microarcsecond astrometry \citep[e.g.,][]{Rioja01.2026.SKA,Curiel01.2026.SKA}.
\item Global baselines, including ground networks, space-VLBI, and partner arrays, to deliver dense $uv$-coverage and high-fidelity imaging \citep[e.g.,][]{Kovalev01.2026.SKA,Bempong-Manful01.2026.SKA,Kadler02.2026.SKA}.
\item Simultaneous multi-scale, multi-frequency imaging, in which the compact structure recovered on VLBI baselines is placed in the context of the extended emission around it, as needed for fast-evolving novae and for complex jet systems such as NGC\,1068 \citep[e.g.,][]{Marcote01.2026.SKA,Panessa01.2026.SKA}. This requires that the SKAO interferometric data taken alongside the tied-array beams be made available to VLBI users, and that observations be configured with multiple sub-arrays to retain short-baseline ($<160$\,km for SKA-Mid and $<75$\,km for SKA-Low) coverage.
\end{itemize}

Together, the chapters outline a broad array of science cases that depend on VLBI -- covering topics from jet origins and particle acceleration physics, to galaxy and black hole formation, the Milky Way's structure, and fundamental physics tests. These cases all require high angular resolution and precise astrometry. Without VLBI-scale baselines alongside the SKAO, many of these scientific goals will remain unattainable.

\section{VLBI with the SKAO: capabilities \& implementation}\label{s:technical}

\begin{figure}
    \centering
    \includegraphics[width=\linewidth]{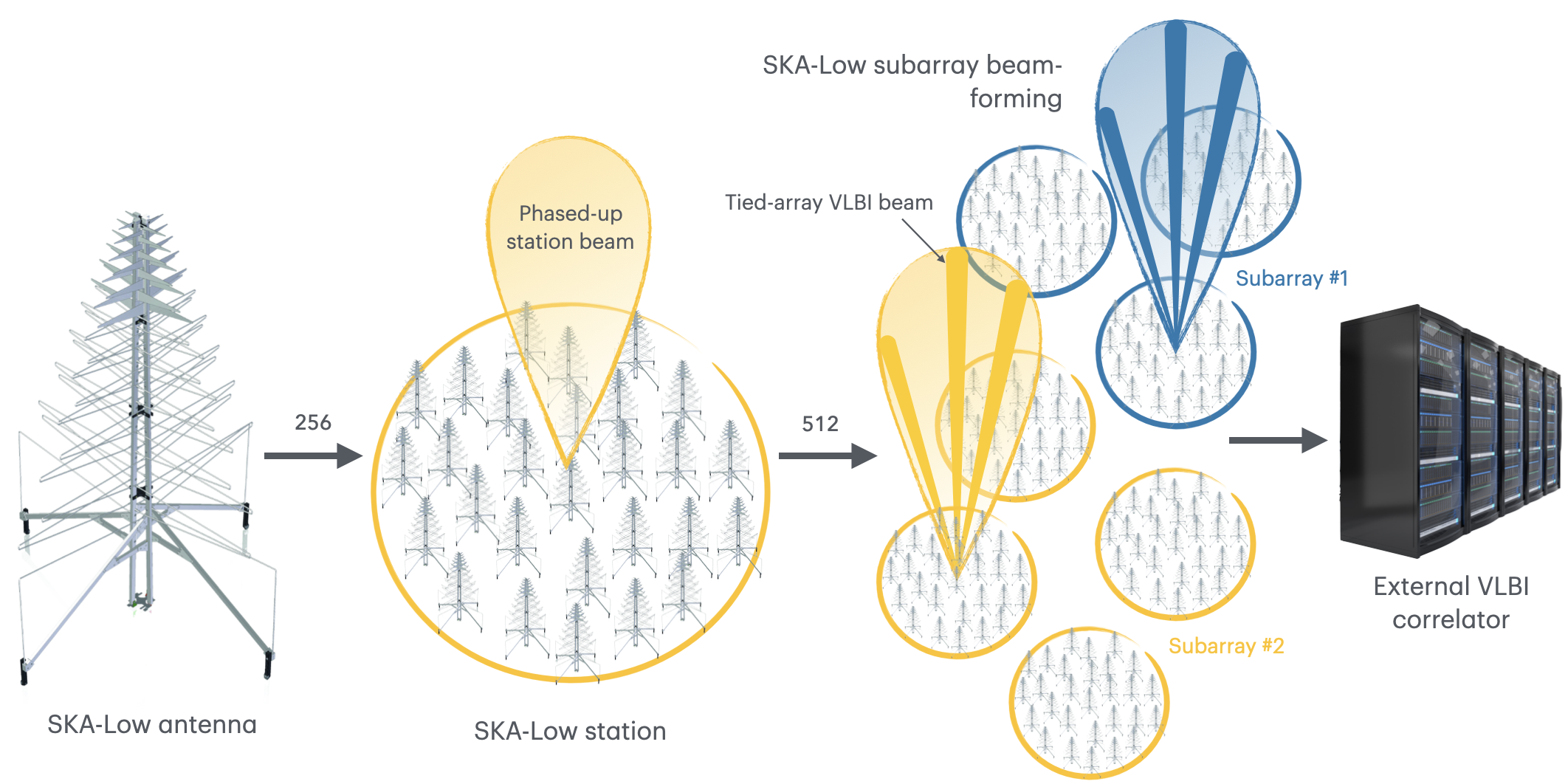}
    \includegraphics[width=\linewidth]{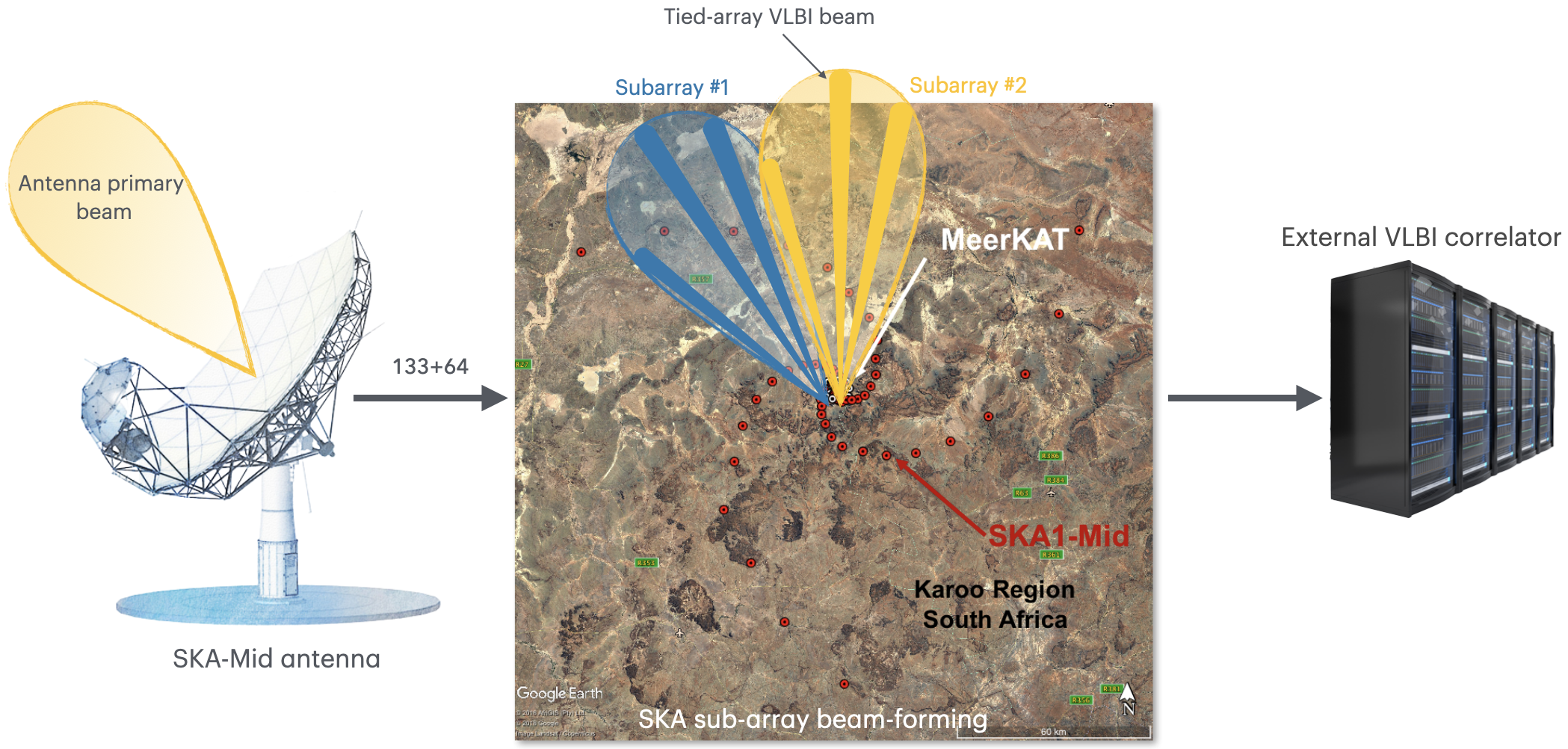}
    \caption{The operational diagram of how SKA-Low (top) and SKA-Mid (bottom) will operate as a VLBI instrument. For both instruments, subarrays will be phased up from a subset of SKA-Low stations (which have 256 phased-up SKA-Low antennas per station) or SKA-Mid dishes to form a tied-array/pencil beam. Each pencil beam will be recorded in a VLBI-compatible format and transferred to an external correlator for combination with other radio telescope data.}
    \label{f:SKA-Mid_VLBI_ops}
\end{figure}

In this section, we summarise the current operational knowledge of how the SKAO will be incorporated into a global VLBI array. Incorporating SKA-Mid and SKA-Low into global VLBI networks requires a dedicated signal path that is distinct from the standard interferometric imaging pipeline. Because the data rates and computational demands of full VLBI correlation are incompatible with routing all SKA antennas through either the SKA correlator or an external VLBI correlator, the SKAO will support VLBI through a tied-array beamforming approach: a sub-array of antennas is phased up into one or more high-sensitivity pencil beams, each of which then serves as a single VLBI ``station'' for correlation with other telescopes in the global network (see Figure~\ref{f:SKA-Mid_VLBI_ops}).

\subsection{Tied-array beamforming and data output}

The Central Signal Processor (CSP) Correlator and Beamformer (CBF) of each telescope first forms tied-array beams from a flexible selection of sub-arrays. For SKA-Mid, up to 16 such beams can be produced simultaneously, and for SKA-Low, up to 8 beams are available. Each beam can, in principle, be drawn from a distinct sub-array of antennas (up to 16 independent sub-arrays for SKA-Mid by Cycle~1), enabling simultaneous multi-field VLBI observations or the commensal use of different sub-arrays for imaging and beamformed science.

The tied-array beam data are then converted into the VLBI Data Interchange Format (VDIF), the international standard for VLBI station data streams. For both SKA-Mid and SKA-Low, this conversion is performed on the Pulsar Timing (PST) beam servers rather than on dedicated CBF hardware. This architectural decision reuses the existing PTUSE-based data acquisition and signal processing software, extended to produce VDIF-formatted output. An analogous approach was adopted for SKA-Mid, enabling early availability of VLBI beams before the AA$^*$ milestone and ahead of the full deployment of CBF-based VLBI firmware. In both cases, the beam allocations are shared between VLBI and pulsar timing (PST) observations, with the allocation between the two modes determined during scheduling. A 100\,Gbps link carries the aggregated VLBI data products from the PST output switch to the wider network. 

\subsection{Data path and external correlation}

The VLBI data stream follows a dedicated signal path that bypasses the standard SKAO data products pipeline and the SKA Regional Centre Network (SRCNet). From the PST beam servers, the VDIF-formatted data are transferred to VLBI recorders at the telescope sites and then transmitted, either in real time ($e$-VLBI) or via recorded media, to an external VLBI correlator for joint processing with the other network stations. The resulting correlated data products are returned to the observer via the operating network's standard VLBI post-processing and imaging workflows.

\subsection{Operational timeline}

The staged delivery of VLBI capability follows the broader SKAO science verification and early-operations schedule. For SKA-Mid, a single VLBI beam is expected to be available during the AA2 Science Verification phase (2029), providing the first opportunity for commissioning the end-to-end VLBI signal chain. By AA$^*$ (2031), this will grow to four simultaneous VLBI beams, and full basic VLBI capability -- with the complete 16-beam pool available across up to 16 sub-arrays -- is planned for Cycle~0 (2032). For SKA-Low, a broadly analogous timeline is anticipated, with early single-beam VLBI capability in the first science verification cycles growing to a potential eight multi-beam operation by Cycle~0. The SKAO \textbf{guarantees} only four simultaneous VLBI beams for SKA-Mid and SKA-Low, and the 16- and 8-beam capability described earlier for SKA-Mid and SKA-Low is on a best-effort basis. These milestones are illustrated at \href{https://www.skao.int/en/science-users/timeline-science}{https://www.skao.int/en/science-users/timeline-science}.

The number of VLBI beams available at any given time is a shared resource with pulsar timing, and the practical allocation will depend on the concurrent observing programme. Several key SKA-VLBI science cases (e.g., multi-field astrometry, multi-source monitoring) require more than four simultaneous beams, and three of the seven identified Low-frequency VLBI science cases need up to eight VLBI beams. The flexibility afforded by the PST-based architecture, where the VLBI and PST beam pools can be dynamically partitioned, is therefore an important operational asset and is expected to be further exploited as the telescope matures.

\section{Challenges and recommendations to the wider community}\label{s:challenges}

\begin{figure}
    \centering
    \includegraphics[width=\linewidth]{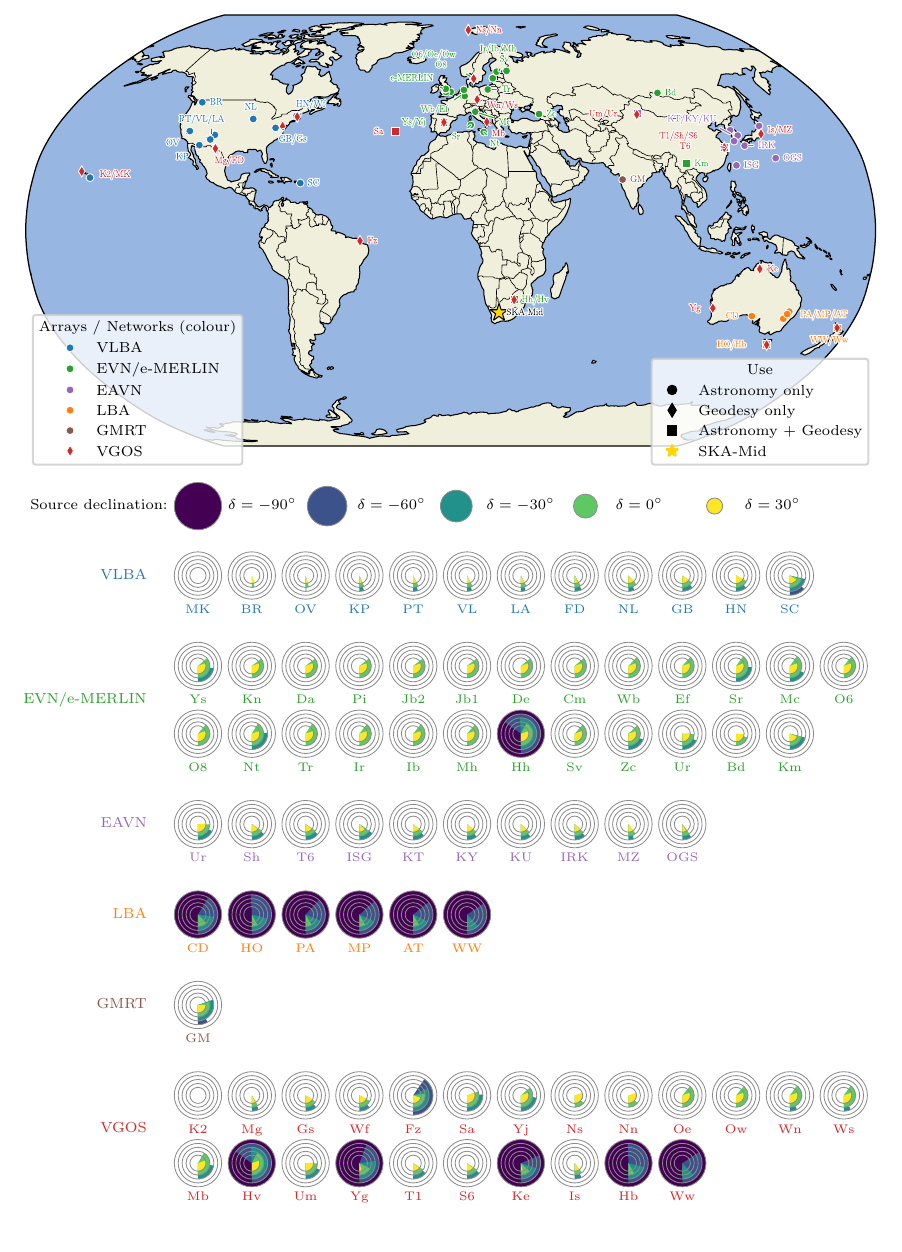}
    \caption{Locations of a selection of radio telescopes with frequency-band overlaps with the SKA-Mid telescope (station acronyms are defined in Appendix~\ref{app:stations}). In the bottom panel, each small circular diagram shows the mutual sky coverage between one telescope (labelled by its acronym) and SKA-Mid. Each diagram contains a set of concentric arcs, one per source declination $\delta$ (the outermost arc is $\delta=-90^\circ$, the innermost $\delta=+30^\circ$), and the colour of an arc indicates that declination (colour scale from $\delta=-90^\circ$ to $+30^\circ$, shown in the key above the grid). The length of each arc is the fraction of a sidereal day during which the two telescopes can simultaneously see a source at that declination, so a full circle means the source is always mutually visible.}
    \label{f:SKA-Mid_VLBI_compatible}
\end{figure}
\begin{figure}
    \centering
    \includegraphics[width=\linewidth]{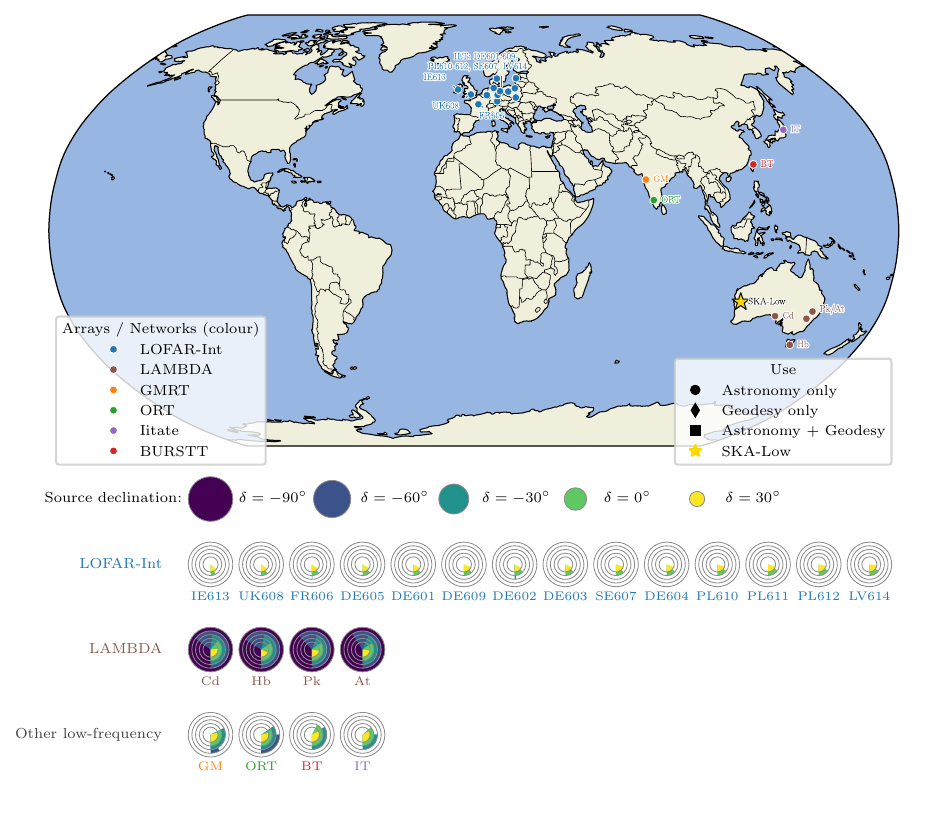}
    \caption{Locations of a selection of radio telescopes with frequency-band overlaps with the SKA-Low telescope (station acronyms are defined in Appendix~\ref{app:stations}). In the bottom panel, each small circular diagram shows the mutual sky coverage between one telescope (labelled by its acronym) and SKA-Low. Each diagram contains a set of concentric arcs, one per source declination $\delta$ (the outermost arc is $\delta=-90^\circ$, the innermost $\delta=+30^\circ$), and the colour of an arc indicates that declination (colour scale from $\delta=-90^\circ$ to $+30^\circ$, shown in the key above the grid). The length of each arc is the fraction of a sidereal day during which the two telescopes can simultaneously see a source at that declination, so a full circle means the source is always mutually visible. Given the smaller number of stations, we have included prospective stations of the LAMBDA telescope \citep[see][]{Timmerman01.2026.SKA}}
    \label{f:SKA-Low_VLBI_compatible}
\end{figure}

In this section, we outline upcoming challenges and open questions for SKA-VLBI. The scientific community will need to participate actively to resolve them. 

Given SKA-VLBI's different operational model and the partnerships with external telescopes and arrays it requires, several of these challenges follow directly from the available array geometry. Figures~\ref{f:SKA-Mid_VLBI_compatible} and~\ref{f:SKA-Low_VLBI_compatible} show the telescopes that could realistically participate in SKA-Mid and SKA-Low VLBI observations, together with the fraction of the sidereal day over which each shares sky with the SKAO as a function of source declination.

\paragraph{Key recommendation: A coordinating body for global VLBI participation in SKAO.} A coordinating body for VLBI needs to be established and is empowered to negotiate policy and provide a single voice to SKAO from the VLBI community. This body would need to include all the key stakeholders for both SKA-Mid and SKA-Low VLBI-capable stations, and would need to consider aspects such as how to transport the data -- given it will not go through SRCNet, where to correlate, how to propose for SKAO VLBI time, and how to provision VLBI data recorders for the SKAO. It would also need to coordinate responsive observing across arrays to support time-domain science cases. 

\paragraph{Utilising the wide SKAO bandwidths and frequency coverage.} The SKAO will provide extremely wide bandwidths ranging from 300 MHz on SKA-Low to 5 GHz on SKA-Mid Band 5b, improving sensitivity, image fidelity, and the ability to characterise the spectral profile of sources. In contrast, current VLBI stations (especially those operating at SKA-Mid frequencies) have limited data rates (e.g., 4\,Gbps for the VLBA, EVN, and LBA; corresponding to a 512\,MHz bandwidth -- assuming 2-bit sampling). While this should be sufficient for the lower-frequency bands, SKA-VLBI will be limited to the bandwidth available from the VLBI arrays. This would be exacerbated if higher bit rates are adopted, for example, to provide greater resilience to radio frequency interference (RFI).

A related limitation applies to the frequency coverage itself. Band 5b extends up to 15.4\,GHz, and observing at the top of this band would deliver the finest angular resolution accessible to SKA-VLBI. However, Ku-band (roughly 12--15\,GHz) receivers are not widely deployed across the stations that SKA-Mid would be correlated against, and Ku-band is not currently a standard observing band for the EVN or the LBA. Without this, the upper part of Band 5b cannot be exploited for VLBI.

\textbf{Recommendations:} VLBI arrays continue to plan for expansion to 16/32 Gbps operational data rates and to develop their data transport, correlation, post-processing and archiving capabilities to handle larger data volumes and increased computational loads. In parallel, stations should consider installing Ku-band receivers, and the EVN and LBA should work towards supporting Ku-band VLBI observations in the future, given that MeerKAT will soon be equipped with these receivers, so that the full frequency range of SKA-Mid Band 5b, and with it the highest possible resolution, becomes available to users. In contrast, the SKAO should consider expanding to cover K-band/Band 6, given the large number of science cases (e.g., \href{https://www.dropbox.com/scl/fi/zi4pwvb3noa66htybsrjo/20-01-Beyond-Band-5.pdf?rlkey=ocoy6rjwl0k0eira8p02fgq22&e=1&dl=0}{SKAO memo 20-01}) it would enable, the higher resolutions that are possible, and that many telescopes in VLBI arrays are already equipped with K-band receivers. 

\begin{figure}
    \centering
    \includegraphics[width=\linewidth]{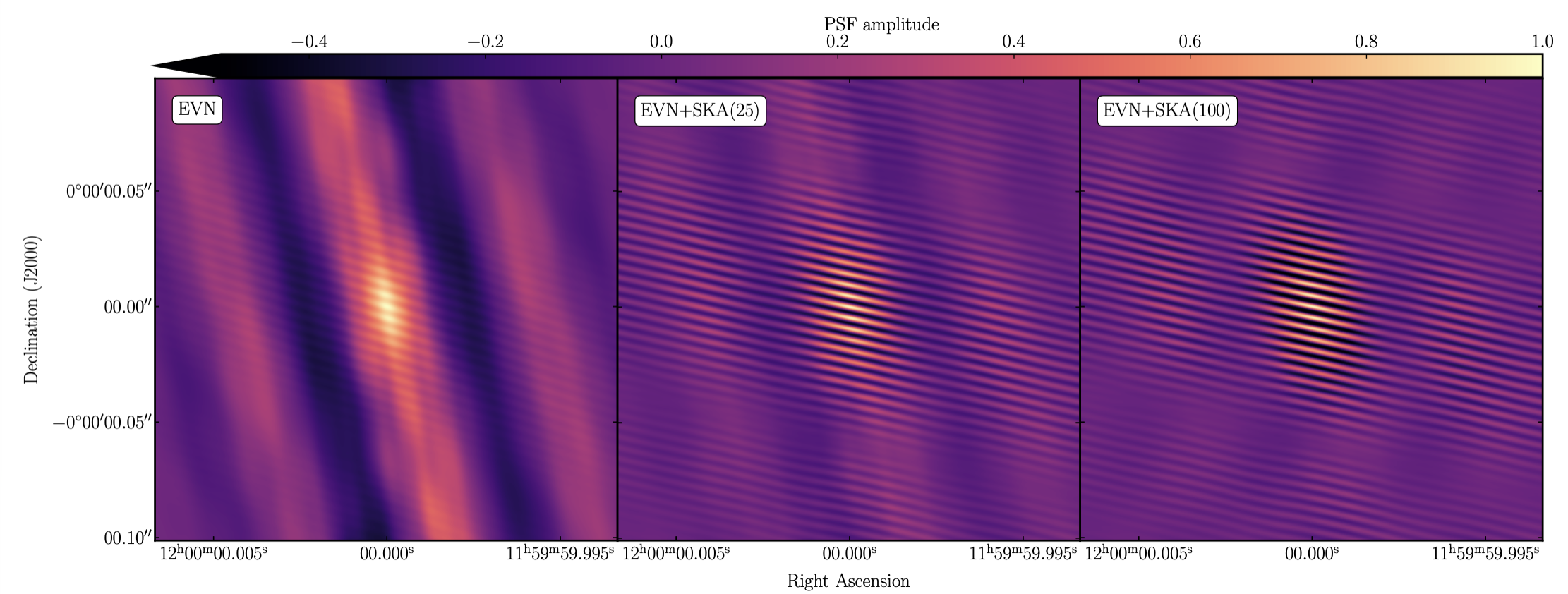}
    \caption{A 12-hour simulated point-spread function of an observation with the European VLBI Network (left panel) plus the inclusion of 25 and 100 phased-up SKA-Mid dishes (central and right panel, respectively) operating at a central frequency of 1.6 GHz. The inclusion of the sensitive SKA-Mid dishes, along with the long North-South baseline orientations of SKA-Mid, degrades the PSF, making accurate deconvolution more challenging.}
    \label{fig:skamid_psf}
\end{figure}

\paragraph{Image fidelity.} The SKA telescopes will provide extremely sensitive single elements to global VLBI arrays (SEFD: $\sim$3.3\,Jy for SKA-Mid AA$^{*}$ at 18 cm, reducing to $\sim$1.55\,Jy for AA4), which will dominate the image weighting when generating images at the best sensitivities. This will result in point spread functions with strong sidelobes and sharp, stripe-like structure, set by the small number of highly weighted baselines to the SKA telescopes, which will make deconvolution and recovery of the true sky brightness distribution difficult (see Figure~\ref{fig:skamid_psf}). This is a similar problem to the one the Event Horizon Telescope has experienced, in which baselines to phased-up ALMA dominate the images.

\noindent\textbf{Recommendations:} Investigate the application of next-generation imaging algorithms \citep[e.g.,][]{2016A&A...586A..76J,2022A&A...666A.137M}, for use in SKA-VLBI. This could be conducted as a new SKA Data Challenge, in which a simulated SKA-Low and SKA-Mid observation is provided to the community, and the goal is to recover the input model as accurately as possible. Adding additional sensitive elements to the participating networks would also help, since a more uniform distribution of baseline orientations across the array reduces sidelobe levels to some degree. Examples include the GMRT at 1.4\,GHz\footnote{\href{https://www.jive.eu/gmrt-evn-vlbi-observations.html}{https://www.jive.eu/gmrt-evn-vlbi-observations.html}} and the upcoming Jingdong 120-m \citep{Chen01.2026.SKA} and Qitai 110-m telescopes.

\paragraph{Increase the number of VLBI partner instruments.} 

For SKA-Mid, \citet{Bempong-Manful01.2026.SKA} highlighted the need for more African VLBI stations to improve image fidelity, especially in equatorial fields, and summarised stations under development, such as the Botswana SKAO telescope and the Africa mm-telescope. As Figure~\ref{f:SKA-Mid_VLBI_compatible} shows, the stations with the longest mutual visibility with SKA-Mid are concentrated in the northern hemisphere, so the north--south baselines that these African stations would provide are precisely the ones the array currently lacks. Further developments in other areas, such as Asia (e.g., the FAST core array, the Thai Radio Telescope, and the Jingdong 120-m and Qitai 110-m telescopes), will further enhance VLBI image quality and sensitivity with the SKAO. Any potential developments in South America could provide key East-West baselines and link the SKAO to the VLBA and ngVLA.

For SKA-Low, \citet{Timmerman01.2026.SKA} have highlighted the wide range of science cases that SKA-Low VLBI can address, including exoplanets, AGN and jet formation, young stellar objects, pulsar astrometry, and more. They also emphasise the unique position of SKA-Low VLBI in the Southern Hemisphere and its ability to target Galactic (e.g., LMC) and extragalactic objects (e.g., Pic A, Cen A) at low frequencies for the first time. However, as shown in Figure~\ref{f:SKA-Low_VLBI_compatible}, there are far fewer stations that can participate in long-baseline SKAO observations than in SKA-Mid. Aside from the ongoing LAMBDA project, very few compatible stations are available or planned before SKA-Low VLBI is operational. This threatens our ability to realise the scientific opportunities offered by SKA-Low. 

\noindent{\textbf{Recommendations:}} The community must actively work to increase the number of SKA-Low and Mid long-baseline instruments. Particular attention should be made to SKA-Low, which has far fewer long-baseline stations, and investment in current initiatives, such as LAMBDA, should be considered. Given current trends in funder and government priorities in developing countries (e.g., Africa, SE-Asia), the community should consider dual-use applications (e.g., Space Situational Awareness, geodesy) and public-private partnerships to support their efforts to secure new infrastructure.


\section{Summary and conclusions}\label{s:conclusions}

 \begin{figure}
     \centering
     \includegraphics[width=1.0\linewidth]{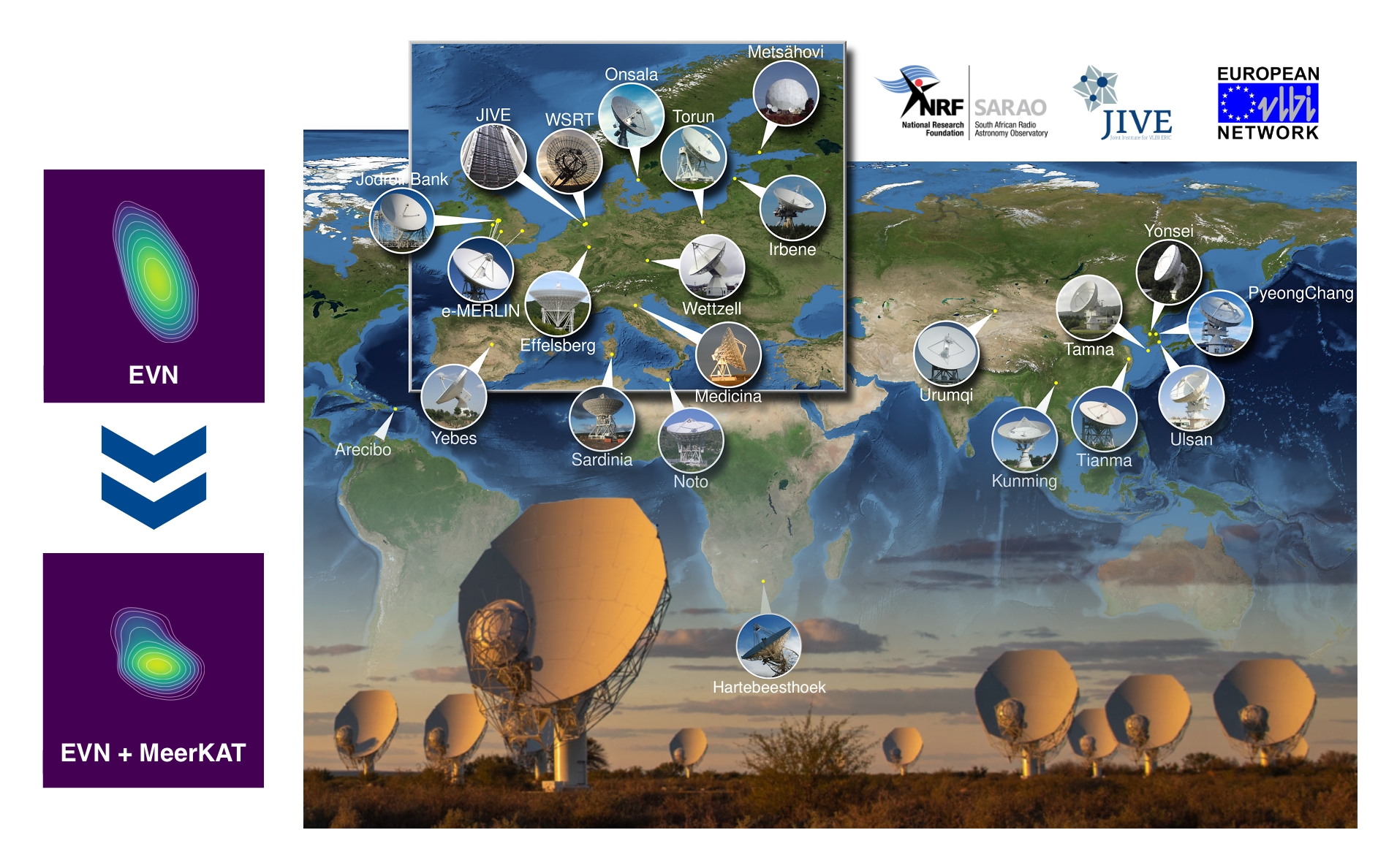}
     \caption{The SKAO precursor MeerKAT observing as a phased-array element of the European VLBI Network. A tied-array beam formed from the MeerKAT dishes is correlated with the other stations of the network, using the same operational model that SKA-Mid will follow. Illustration: Benito Marcote, JIVE. Image credits: MeerKAT image $\copyright$SARAO; EVN map image $\copyright$Paul Boven, JIVE (satellite image: Blue Marble Next Generation, courtesy of NASA Visible Earth).}
     \label{f:jive_sarao_vlbi_pr}
 \end{figure}

In summary, Very Long Baseline Interferometry (VLBI) will be a vital mode for the SKAO, enabling high-resolution and high-sensitivity observations that the SKAO alone cannot perform (see Figure~
\ref{fig:las_skavlbi}). As discussed in Section~\ref{s:science}, SKAO’s VLBI capability is key to many scientific applications, including the study of jets in AGN and X-ray binaries, microsecond astrometry for defining the celestial reference frame and galactic structure, pulsar parallax measurements for timing, and tests of general relativity.

Furthermore, integrating SKAO into VLBI arrays is now quite well understood. According to Section~\ref{s:technical}, both telescopes will generate tied-array beams from sub-arrays via the PST system within the SKAO. This will happen in stages, offering up to 16 tied-array beams for SKA-Mid and 8 for SKA-Low. These beams will be saved in a VLBI-compatible format and sent to an external facility for correlation with data from other VLBI telescopes. This approach has been demonstrated with SKAO precursors, such as MeerKAT, which has been phased-up and correlated successfully with telescopes from the European VLBI Network (see Figure~\ref{f:jive_sarao_vlbi_pr}) and the Southern Hemisphere Long Baseline Array. These experiments serve as proof of concept for SKAO’s potential operation with VLBI arrays.

Since VLBI involves collaboration among multiple telescopes worldwide, the scientific benefits of SKA-VLBI depend on the arrays it partners with. Accordingly, recommendations from Section~\ref{s:challenges} are based on this. VLBI arrays should continue upgrading to 16/32 Gbps data rates to utilise SKAO bandwidths and work toward Ku-band support to access higher parts of Band 5b. Developing and testing imaging and deconvolution techniques that can handle the non-uniform weighting introduced by SKA telescopes is essential, and a dedicated SKA Data Challenge would be suitable for this. The number of long-baseline partner stations, especially for SKA-Low, needs to increase, as currently few compatible stations are available or planned. Lastly, a coordinating organisation is necessary to represent the VLBI community to SKAO collectively, addressing issues like data transport, correlation, proposals, recording facilities, and organising quick-response observing, which are beyond the SKAO’s direct scope. While not insurmountable, each step requires community effort rather than just the SKAO alone.

\clearpage
\appendix
\section{VLBI and geodetic station acronyms}
\label{app:stations}

Tables~\ref{tab:stations-mid} and~\ref{tab:stations-low} give the full names and locations of the stations and networks identified by acronym in Figures~\ref{f:SKA-Mid_VLBI_compatible} and~\ref{f:SKA-Low_VLBI_compatible}, respectively. Coordinates, diameters and other technical details for these stations are also maintained by their respective networks and consortia (e.g., the IVS network station list\footnote{\url{https://ivscc.gsfc.nasa.gov/stations/}}, the EVN\footnote{\url{https://www.evlbi.org/stations}}, the VLBA\footnote{\url{https://public.nrao.edu/telescopes/vlba/}}, and the LBA\footnote{\url{https://www.atnf.csiro.au/vlbi/overview/}}).

{\small
\begin{longtable}{ll}
\caption{Stations shown in Figure~\ref{f:SKA-Mid_VLBI_compatible}, grouped by network.}
\label{tab:stations-mid} \\
\hline
Acronym & Station \\
\hline
\endfirsthead
\multicolumn{2}{l}{\tablename~\thetable{} -- continued} \\
\hline
Acronym & Station \\
\hline
\endhead
\hline
\endfoot
\multicolumn{2}{l}{\textit{VLBA (and associated High Sensitivity Array antennas)}} \\
BR & Brewster, WA, USA \\
FD & Fort Davis, TX, USA \\
HN & Hancock, NH, USA \\
KP & Kitt Peak, AZ, USA \\
LA & Los Alamos, NM, USA \\
NL & North Liberty, IA, USA \\
OV & Owens Valley, CA, USA \\
PT & Pie Town, NM, USA \\
SC & Saint Croix, US Virgin Islands \\
MK & Mauna Kea, HI, USA \\
GB & Green Bank Telescope (100\,m), WV, USA \\
VL & Karl G. Jansky Very Large Array (phased), Socorro, NM, USA \\
\hline
\multicolumn{2}{l}{\textit{EVN / e-MERLIN}} \\
Ef & Effelsberg (100\,m), Germany \\
Sr & Sardinia Radio Telescope (64\,m), Italy \\
Ys & Yebes (40\,m), Spain \\
Mc & Medicina (32\,m), Italy \\
Nt & Noto (32\,m), Italy \\
Tr & Toru\'n (32\,m), Poland \\
O6 & Onsala (20\,m), Sweden \\
O8 & Onsala (25\,m), Sweden \\
Ir & Irbene (32\,m), Latvia \\
Ib & Irbene (16\,m), Latvia \\
Wb & Westerbork (25\,m, single dish), the Netherlands \\
Jb2 & Jodrell Bank Mk~II (32\,m), UK \\
Jb1 & Lovell Telescope (76\,m), Jodrell Bank, UK \\
Cm & Cambridge (32\,m), UK \\
Ur & Urumqi (25\,m), China \\
Km & Kunming (40\,m), China \\
Bd & Badary (32\,m), Russia \\
Zc & Zelenchukskaya (32\,m), Russia \\
Sv & Svetloe (32\,m), Russia \\
Mh & Mets\"ahovi (14\,m), Finland \\
Hh & Hartebeesthoek (26\,m), South Africa \\
Kn & e-MERLIN Knockin (25\,m), UK \\
De & e-MERLIN Defford (25\,m), UK \\
Da & e-MERLIN Darnhall (25\,m), UK \\
Pi & e-MERLIN Pickmere (25\,m), UK \\
\multicolumn{2}{l}{\hspace{1em}(T6: Tianma and Sh: Sheshan, listed below under EAVN, are also EVN stations)} \\
\hline
\multicolumn{2}{l}{\textit{EAVN (East Asian VLBI Network)}} \\
KY & KVN Yonsei (21\,m), South Korea \\
KU & KVN Ulsan (21\,m), South Korea \\
KT & KVN Tamna, Jeju (21\,m), South Korea \\
MZ & VERA Mizusawa (20\,m), Japan \\
IRK & VERA Iriki (20\,m), Japan \\
ISG & VERA Ishigakijima (20\,m), Japan \\
OGS & VERA Ogasawara (20\,m), Japan \\
T6 & Tianma (65\,m), China \\
Sh & Sheshan (25\,m), China \\
\multicolumn{2}{l}{\hspace{1em}(Ur: Urumqi, listed above under EVN, also participates in EAVN)} \\
\hline
\multicolumn{2}{l}{\textit{LBA (Australian Long Baseline Array)}} \\
AT & ATCA (phased array), Narrabri, Australia \\
PA & Parkes (64\,m), Australia \\
MP & Mopra (22\,m), Australia \\
HO & Hobart (26\,m), Australia \\
CD & Ceduna (30\,m), Australia \\
WW & Warkworth (12\,m), New Zealand \\
\hline
\multicolumn{2}{l}{\textit{GMRT}} \\
GM & Giant Metrewave Radio Telescope (phased array), Khodad, India \\
\hline
\multicolumn{2}{l}{\textit{VGOS (VLBI Global Observing System, geodetic)}} \\
Wn & Wettzell twin telescope 1, Germany \\
Ws & Wettzell twin telescope 2, Germany \\
Wf & Westford (18.3\,m), MA, USA \\
K2 & Kokee Park (12\,m), HI, USA \\
Gs & GGAO12, Greenbelt, MD, USA \\
Oe & Onsala twin telescope 1, Sweden \\
Ow & Onsala twin telescope 2, Sweden \\
Hb & Hobart (12\,m), Australia \\
Ke & Katherine (12\,m), Australia \\
Yg & Yarragadee (12\,m), Australia \\
Is & Ishioka (13\,m), Japan \\
Nn & Ny-\AA lesund twin telescope 1, Norway \\
Ns & Ny-\AA lesund twin telescope 2, Norway \\
Hv & Hartebeesthoek VGOS (13.2\,m), South Africa \\
Um & Urumqi VGOS (13\,m), China \\
Mb & Matera VGOS (13.2\,m), Italy \\
Yj & Yebes VGOS (13.2\,m), Spain \\
Sa & Santa Maria (13.2\,m), Azores, Portugal \\
S6 & Sheshan VGOS (13\,m), China \\
T1 & Tianma VGOS (13.2\,m), China \\
Mg & McDonald (12\,m), TX, USA \\
Fz & Fortaleza VGOS (12\,m), Brazil \\
\multicolumn{2}{l}{\hspace{1em}(Ww: Warkworth 12\,m, listed above under LBA as WW, is also a VGOS station)} \\
\hline
\end{longtable}
}

{\small
\begin{longtable}{ll}
\caption{Stations shown in Figure~\ref{f:SKA-Low_VLBI_compatible}, grouped by network.}
\label{tab:stations-low} \\
\hline
Acronym & Station \\
\hline
\endfirsthead
\multicolumn{2}{l}{\tablename~\thetable{} -- continued} \\
\hline
Acronym & Station \\
\hline
\endhead
\hline
\endfoot
\multicolumn{2}{l}{\textit{LOFAR international stations}} \\
IE613 & Birr, Ireland \\
UK608 & Chilbolton, UK \\
FR606 & Nan\c{c}ay, France \\
DE601 & Effelsberg, Germany \\
DE602 & Unterweilenbach, Germany \\
DE603 & Tautenburg, Germany \\
DE604 & Bornim (Potsdam), Germany \\
DE605 & J\"ulich, Germany \\
DE609 & Norderstedt (Hamburg), Germany \\
SE607 & Onsala, Sweden \\
PL610 & Bor\'owiec (Pozna\'n), Poland \\
PL611 & Baldy, Poland \\
PL612 & \L azy, Poland \\
LV614 & Irbene, Latvia \\
\hline
\multicolumn{2}{l}{\textit{LAMBDA (prospective low-frequency stations at existing LBA sites)}} \\
At & Narrabri (ATCA site), Australia \\
Pk & Parkes, Australia \\
Cd & Ceduna, Australia \\
Hb & Hobart, Australia \\
\hline
\multicolumn{2}{l}{\textit{Other low-frequency telescopes}} \\
GM & GMRT, Khodad, India \\
ORT & Ooty Radio Telescope, India \\
IT & Iitate Observatory (Tohoku University), Japan \\
BT & BURSTT, Fushan, Taiwan \\
\hline
\end{longtable}
}

\bibliographystyle{abbrvnat-maxbibnames4}
\bibliography{chapter} 

\clearpage
\section{Full author list and affiliations}
\label{app:fullauthorlist}

The complete list of co-authors of this chapter is given below, followed by their affiliations. Authors are listed alphabetically, except for the lead authors, who appear first, as on the title page.

{\raggedright\large
\ExplSyntaxOn\seq_clear:N\l_authors_s\seq_clear:N\l_super_s\ExplSyntaxOff
\author[1,2,3]{Jack~F.~Radcliffe\orcidlink{0000-0002-0813-0497}}
\author[4]{Jun~Yang}
\author[5]{Zsolt~Paragi}
\author[6]{Olga~Bayandina\orcidlink{https://orcid.org/0000-0003-4116-4426}}
\author[3]{Cormac~Reynolds}
\author[8]{Cristina~García-Miró}
\author[7]{Tao~An}
\author[9]{M.~Abada-Simon}
\author[60]{J.N.H.S.~Aditya}
\author[11]{Iv\'an~Agudo}
% ...
\printAuthors
\par}

\affiliation[9]{LESIA, Observatoire de Paris, Universit\'e PSL, CNRS, Sorbonne Universit\'e, Universit\'e de Paris, 5 place Jules Janssen, 92195 Meudon, France}
% ...
\affiliation[11]{Instituto de Astrof\'isica de Andaluc\'ia -- CSIC, Glorieta de la Astronom\'ia, E-18008, Granada, Spain}
% ...
\affiliation[60]{National Centre for Radio Astrophysics - Tata Institute of Fundamental Research, S. P. Pune University Campus, Pune 411007, India}
% ...
\affiliation[80]{(1) Joint Institute for VLBI ERIC, Oude Hoogeveensedijk 4, 7991 PD Dwingeloo, The Netherlands
(2) ASTRON, The Netherlands Institute for Radio Astronomy, Oude Hoogeveensedijk 4, 7991 PD Dwingeloo, The Netherlands} 
% ...
\affiliation[83]{(1) Centre for Extragalactic Astronomy, Department of Physics, Durham University, South Road, Durham DH1 3LE, UK
(2) Institute for Computational Cosmology, Department of Physics, Durham University, South Road, Durham DH1 3LE, UK} 
% ...
\affiliation[114]{(1) Centre for Extragalactic Astronomy, Department of Physics, Durham University, Durham DH1 3LE, UK
(2) Institute for Computational Cosmology, Department of Physics, Durham University, South Road, Durham DH1 3LE, UK} 
\affiliation[115]{(1) Dipartimento di Fisica e Astronomia, Università di Bologna, via Gobetti 93/2, I-40129 Bologna, Italy
(2) Istituto Nazionale di Astrofisica - Istituto di Radioastronomia (IRA), via Gobetti 101, I-40129 Bologna, Italy} 
% ...

\begin{list}{}{%
  \setlength{\leftmargin}{0.2cm}%
  \setlength{\labelsep}{0pt}%
  \setlength{\labelwidth}{1em}%
  \setlength{\itemsep}{\affiliationsSep}%
  \setlength{\topsep}{-0.5\parskip}}
\itshape\small
\makeatletter\the\affil@toks\makeatother
\end{list}

\end{document}